\documentclass[draftcls,journal, twoside, onecolumn,10pt]{IEEEtran}
\usepackage{amsmath}
\usepackage{amssymb}

\newcommand{\Arikan}{Ar\i kan}
\newcommand{\Sasoglu}{\c{S}a\c{s}o\u{g}lu}
\newcommand{\prob}{\mathrm{Pr}}

\newtheorem{lemm}{Lemma}

\newcommand{\bibfilePath}{../../../ee/mybib}
\newcommand{\twobibs}[2]{#2} 

\begin{document}
\title{A Simple Proof of Fast Polarization}
\author{Ido Tal\\
Department of Electrical Engineering \\
Technion -- Haifa 32000, Israel\\
E-mail: idotal@ee.technion.ac.il

}
\maketitle
\begin{abstract}
Fast polarization is an important and useful property of polar codes. It was proved for the binary polarizing $2 \times 2$ kernel by \Arikan\ and Telatar. The proof was later generalized by \Sasoglu. We give a simplified proof.
\end{abstract}


\section{Introduction}
Polar codes are a novel family of error correcting codes, invented by \Arikan\ \cite{Arikan:09p}. The seminal definitions and assumptions in \cite{Arikan:09p} were soon expanded and generalized. Key to almost all the results involving polar codes is the concept of fast polarization. The essence of fast polarization is the phenomenon stated in the following lemma. The lemma was used implicitly by Korada, \Sasoglu, and Urbanke \cite[proof of Theorem 11]{KSU:10p}, and is a generalization of a result by \Arikan\ and Telatar \cite[Theorem 3]{ArikanTelatar:09c}. Its explicit formulation and full proof are due to \Sasoglu\ \cite[Lemma 5.9]{Sasoglu:12b}.
\begin{lemm}
\label{lemm:eren}
Let $B_0, B_1, \ldots$ be an i.i.d.\ process where $B_0$ is uniformly distributed over $\{1, 2, \ldots, \ell\}$. Let $Z_0, Z_1, \ldots$ be a $[0, 1]$-valued random process where 
\begin{equation}
\label{eq:zmplusone}
Z_{m+1} \leq K \cdot Z_m^{D_t} \; , \quad \mbox{whenever $B_m = t$} \; .
\end{equation}
We assume $K \geq 1$ and $D_1, D_2, \ldots , D_\ell > 0$. Suppose also
that $Z_m$ converges almost surely to a $\{0, 1\}$-valued random variable $Z_\infty$. Then, for any
\[
0 < \beta < E \triangleq \frac{1}{\ell} \sum_{t=1}^\ell \log_\ell D_t  
\]
we have
\begin{equation}
\label{eq:lemm:eren}
\lim_{m \to \infty} \prob[Z_m \leq 2^{-\ell^{\beta \cdot m}} ] = \prob[Z_\infty = 0] \; .
\end{equation}
\end{lemm}

The lemma is used to prove that the Bhattacharyya parameter associated with a random variable that underwent polarization (for example, a synthesized channel) polarizes to $0$ at a rate faster than polynomial \cite[Theorem 5.4]{Sasoglu:12b}. A similar claim holds in the case of polarization of the Bhattacharyya parameter to $1$ \cite[Theorem 16]{KoradaUrbanke:10p}. 

The original proof \cite[Lemma 5.9]{Sasoglu:12b} of Lemma~\ref{lemm:eren} is somewhat involved. To summarize, if $K$ were equal to $1$, the proof would follow almost directly from the strong law of large numbers. However, for $K>1$, a sequence of bootstrapping arguments is applied. That is, a current bound on the rate of convergence of $Z_m$ to $0$ is used to derive a stronger bound, and the process is repeated.

The main aim of this paper is to give a simpler proof of Lemma~\ref{lemm:eren}. Thus, we hopefully give insight into the simple mechanics that are at play. Our simpler proof also leads to a stronger result. That is, we will prove the following, which implies Lemma~\ref{lemm:eren}.
\begin{lemm}
\label{lemm:almostSure}
Let $\{B_m\}_{m=0}^\infty$, $\{Z_m\}_{m=0}^\infty$, $K$, and $E$ be as in Lemma~\ref{lemm:eren}. Then, for $0 < \beta < E$,
\begin{equation}
\label{eq:lemm:almostSure}
\lim_{m_0 \to \infty} \prob[\mbox{$Z_{m} \leq 2^{-\ell^{\beta \cdot m}}$ for all $m \geq m_0$} ] = \prob[Z_\infty = 0] \; .
\end{equation}
\end{lemm}

Note that Lemma~\ref{lemm:almostSure} has an ``almost sure flavor'' \cite[page 69, Equation (2)]{Chung:01b}, while Lemma~\ref{lemm:eren} has an ``in probability flavor'' \cite[page 70, Equation (5)]{Chung:01b}. We prove Lemma~\ref{lemm:almostSure} in Section \ref{sec:lemm:almostSure} and show that it implies Lemma~\ref{lemm:eren} in Section \ref{sec:lemm:eren}.

\section{Proof of Lemma~\ref{lemm:almostSure}}
\label{sec:lemm:almostSure}
Let $\epsilon_a,\epsilon_b > 0$ and $m_a < m_b$ be given constants, specified towards the end. We now define three events, denoted $A$, $B$, and $C$.
\begin{IEEEeqnarray}{ll}
A:\qquad & |Z_m - Z_\infty| \leq \epsilon_a \; , \quad \mbox{for all $m \geq m_a$} \; . \\
B:& \left|\frac{\left| \left\{ m_a \leq i < m : B_i = t\right\}\right|}{m-m_a}  - \frac{1}{\ell}\right| \leq \epsilon_b \; , \quad \mbox{for all $m \geq m_b$ and $1 \leq t \leq \ell$} \; . \\
C:& Z_\infty = 0 \; .
\end{IEEEeqnarray}

Recall that the $Z_m$ converge almost surely to $Z_\infty$. Thus, essentially by definition (see \cite[Theorem 4.1.1.]{Chung:01b}), we have for any fixed $\epsilon_a > 0$ that
\[
\lim_{m_a \to \infty} \prob[A] = 1 \; .
\]
Note that event $B$ is concerned with the frequency of $t$ in the subsequence of i.i.d. random variables $B_{m_a},B_{m_a+1},\ldots,B_{m-1}$, which are uniform over $\{1,2,\ldots,\ell\}$. Thus, by the strong law of large numbers\footnote{The strong law of large numbers is applied $\ell$ times. Each application is with respect to the indicators $B_i = t$, where $1 \leq t \leq \ell$. As before, we use \cite[Theorem 4.1.1.]{Chung:01b}.} \cite[Theorem 5.4.2.]{Chung:01b}, we have for any fixed $\epsilon_b$ and $m_a$ that
\[
\lim_{m_b \to \infty} \prob[B] = 1 \; .
\]

We deduce that for any $\delta_a,\delta_b > 0$ there exist $m_a < m_b$ such that
\begin{equation}
\label{eq:probA}
\prob[A] \geq 1- \delta_a  
\end{equation}
and
\begin{equation}
\label{eq:probB}
\prob[B] \geq 1- \delta_b \; .
\end{equation}
Hence,
\begin{equation}
\label{eq:intersectionBound}
\prob[A \cap B \cap C] \geq \prob[Z_\infty = 0] - \delta_a - \delta_b \; .
\end{equation}

Let us see what the event $A \cap B \cap C$ implies. For fixed $0 < \epsilon_a, \epsilon_b, \delta_a, \delta_b < 1$, let $m_a$ and $m_b$ be as above. Define the shorthand
\[
\theta \triangleq - \log_{\epsilon_a} K \; .
\]
Note that $\theta$ is non-negative, and approaches $0$ as $\epsilon_a$ approaches $0$.
By the definition of the events $A$ and $C$, we have that $Z_m \leq \epsilon_a$ when $m \geq m_a$. Thus, $K \leq Z_m^{-\theta}$ when $m \geq m_a$. Hence, we simplify (\ref{eq:zmplusone}) to
\begin{equation}
\label{eq:simplezmplusone}
Z_{m+1} \leq Z_m ^{D_t - \theta} \; , \quad \mbox{whenever $m \geq m_a$ and $B_m = t$} \; .
\end{equation}

The above equation is the heart of the proof: we have effectively managed to ``make $K$ equal $1$'' --- the simple case discussed earlier. We have ``paid'' for this simplification by having the exponents be $D_t - \theta$ instead of the original $D_t$. However, since $\theta$ can be made arbitrarily close to $0$, this will not be a problem. Essentially, all that remains is some simple algebra, followed by taking the relevant parameters small/large enough. We do this now.

Let us assume $\epsilon_a$ is small enough such that $D_t - \theta > 0$ for all $1 \leq t \leq \ell$ and that $\epsilon_b < 1/\ell$. Recall also that $Z_{m_a} \in [0,1]$. Combining (\ref{eq:simplezmplusone}) with event $B$, we deduce that for all $m \geq m_b$,
\begin{equation}
\label{eq:simplezm}
Z_m \leq Z_{m_a} ^{\prod_{t=1}^\ell (D_t - \theta)^{(m-m_a) \cdot (1/\ell \pm \epsilon_b)}}\; ,
\end{equation}
where the above ``$\pm$'' notation is in fact a function of $t$, defined as
\[
\pm \triangleq 
\begin{cases}
+ & \mbox{if $D_t - \theta \leq 1$,} \\ 
- & \mbox{otherwise.}
\end{cases}
\]
By the definition of event $A$, we have that $Z_{m_a} \leq \epsilon_a$. We will further assume that $\epsilon_a \leq 1/2$. Hence, (\ref{eq:simplezm}) simplifies to the claim that for all $m \geq m_b$,
\begin{equation}
\label{eq:zmwithdelta}
Z_m \leq 2^{-\prod_{t=1}^\ell (D_t - \theta)^{(m-m_a) \cdot (1/\ell \pm \epsilon_b)}} = 2^{- \ell^{(E-\Delta)m}} \; ,
\end{equation}
where
\begin{equation}
\label{eq:Delta}
\Delta =  \sum_{t=1}^\ell \frac{1}{\ell} \log_\ell\left( \frac{D_t}{D_t-\theta} \right) - \sum_{t=1}^\ell \pm \epsilon_b \log_\ell(D_t - \theta) +  \sum_{t=1}^\ell \frac{m_a}{m} \left( \frac{1}{\ell} \pm \epsilon_b \right) \log_\ell(D_t - \theta) \; .
\end{equation}

In light of (\ref{eq:lemm:almostSure}), our goal is to show that for a given $\beta < E$ and $\delta_a,\delta_b>0$,  we can choose $m_a < m_b$, and $\epsilon_a,\epsilon_b > 0$ as above such that $\Delta < E-\beta$. We do this by showing that each of the three sums in (\ref{eq:Delta}) can be made smaller than $(E-\beta)/3$. Recalling that $\theta$ goes to $0$ as $\epsilon_a$ tends to $0$, we deduce that the first sum can be made smaller than $(E-\beta)/3$ by taking $\epsilon_a$ small enough. Similarly, we can make the second sum smaller than $(E-\beta)/3$ by taking $\epsilon_b$ small enough. For the third sum, we first fix $m_a$ large enough such that (\ref{eq:probA}) holds (note that event $A$ is a function of $\epsilon_a$, which is by now fixed). Lastly, we take $m_b$ large enough such that the third sum is smaller than $(E-\beta)/3$ for all $m \geq m_b$, and (\ref{eq:probB}) holds (again, note that event $B$ is a function of  $m_a$ and $\epsilon_b$, which have been fixed).

Recall that our aim is to prove (\ref{eq:lemm:almostSure}). We deduce from (\ref{eq:intersectionBound}), (\ref{eq:zmwithdelta}), and the above paragraph that for all $\delta_a,\delta_b > 0$ and $0 < \beta < E$,
\[
\lim_{m_0 \to \infty} \prob[\overbrace{\mbox{$Z_{m} \leq 2^{-\ell^{\beta \cdot m}}$ for all $m \geq m_0$}}^D ] \geq \prob[Z_\infty = 0] - \delta_a - \delta_b \; .
\]
Indeed, we have just proved that for the parameters fixed as above, $A \cap B \cap C$ implies $D$, for $m_0 = m_b$. Since we are taking the limit of a strictly increasing sequence, the assertion follows (and the limit exists, since the sequence is bounded).

Since the above inequality holds for all $\delta_a,\delta_b > 0$, it must also hold for $\delta_a = \delta_b = 0$. Thus, all that remains is to prove that
\[
\lim_{m_0 \to \infty} \prob[\mbox{$Z_{m} \leq 2^{-\ell^{\beta \cdot m}}$ for all $m \geq m_0$} ] \leq \prob[Z_\infty = 0] \; .
\]
Assume to the contrary that there exists $0 < \beta < E$ and $m_0$ such that
\[
\prob[\mbox{$Z_{m} \leq 2^{-\ell^{\beta \cdot m}}$ for all $m \geq m_0$} ] > \prob[Z_\infty = 0] \; .
\]
Clearly, this implies that
\[
\prob[\lim_{m \to \infty} Z_{m} = 0] > \prob[Z_\infty = 0] \; .
\]
Hence,
\[
\prob[\lim_{m \to \infty} Z_{m} = Z_\infty] < 1 \; ,
\]
contradicting that fact that the sequence $Z_m$ converges almost surely to $Z_\infty$.

\section{Proof of Lemma~\ref{lemm:eren}}
\label{sec:lemm:eren}
We now explain why Lemma~\ref{lemm:almostSure} implies Lemma~\ref{lemm:eren}. That is, why (\ref{eq:lemm:almostSure}) implies (\ref{eq:lemm:eren}). Clearly, (\ref{eq:lemm:almostSure}) implies
\[
\liminf_{m \to \infty} \prob[Z_m \leq 2^{-\ell^{\beta \cdot m}} ] \geq  \prob[Z_\infty = 0] \; .
\]
Thus, the claim will follow if we prove that
\[
\limsup_{m \to \infty} \prob[Z_m \leq 2^{-\ell^{\beta \cdot m}} ] \leq  \prob[Z_\infty = 0] \; .
\]
Assume to the contrary that there exists $0 < \beta < E$ such that
\[
\limsup_{m \to \infty} \prob[Z_m \leq 2^{-\ell^{\beta \cdot m}} ] >  \prob[Z_\infty = 0] \; .
\]
The above implies that the $Z_m$ cannot converge in probability to $Z_\infty$ \cite[page 70, Equation (5)]{Chung:01b}. This contradicts the fact that almost sure convergence implies convergence in probability \cite[Theorem 4.2]{Chung:01b}.


\twobibs{
\bibliographystyle{IEEEtran}
\bibliography{\bibfilePath}
}
{
\ifdefined\bibstar\else\newcommand{\bibstar}[1]{}\fi

}

\end{document}